# Photoinduced Interlayer Dynamics in $T_d$-MoTe$_2$: A Broadband Pump-Probe Study

Meixin Cheng[a], Shazhou Zhong[b], Nicolas Rivas[a#], Tina Dekker[b], Ariel Alcides Petruk[a], Patrick Gicala[a], Kostyantyn Pichugin[a], Fangchu Chen[b], Xuan Luo[c], Yuping Sun[c,d,e], Adam W. Tsen[b], Germán Sciaini[a*].

[*]Correspondence: gsciaini@uwaterloo.ca

[#]Present address: Nuclear Engineering Group, McMaster University, Hamilton, Ontario, L8S 4K1, Canada.

a. The Ultrafast electron Imaging Lab, Department of Chemistry, and Waterloo Institute for Nanotechnology, University of Waterloo, Waterloo, Ontario N2L 3G1, Canada

b. Institute for Quantum Computing, Department of Physics and Astronomy, Department of Electrical and Computer Engineering, and Department of Chemistry, University of Waterloo, Waterloo, Ontario N2L 3G1, Canada

c. Key Laboratory of Materials Physics, Institute of Solid-State Physics, Chinese Academy of Sciences, Hefei 230031, People's Republic of China

d. High Magnetic Field Laboratory, Chinese Academy of Sciences, Hefei, 230031, China

e. Collaborative Innovation Center of Advanced Microstructures, Nanjing University, Nanjing, 210093, China

**Abstract**

We report on time-resolved broadband transient reflectivity (tr-bb-TR) measurements performed on a bulk single crystal of $T_d$-MoTe$_2$ as a function of the incident pump fluence ($F$). Tr-bb-TR data unveil photoinduced electronic changes progressing on the sub-picosecond timescale as well as the dynamics of the coherent low-frequency $^1A_1$ interlayer shear phonon. Our results indicate a gradual evolution of both the TR and the $^1A_1$ Fourier intensity spectra as a function of $F$, ruling out the threshold-like change that has been associated with the ultrafast photoinduced $T_d \rightarrow 1T'$ phase transition. We also observe a large redshift of the $^1A_1$ Fourier spectral features, which suggests that large renormalization effects are taking place on interband transitions that are dielectrically susceptible to the $^1A_1$ interlayer phonon displacement.

MoTe$_2$ is one of the few two-dimensional transition metal dichalcogenides (2D-TMDCs) that presents different polytypes with phase transitions that can be further controlled by temperature[1], alloying[2], strain[3], electrostatic gating[4], and dimensionality[5]. Relevant to the present work is the non-centrosymmetric $T_d$ state[6–8], which is purported to be a Weyl semimetallic phase[6–8] and displays an orthorhombic unit cell with a $c$-axis inclination angle ($\alpha$) of 90° with respect to the plane of the layer. Above a critical temperature of $T_c \approx 250$ K, $T_d$-MoTe$_2$ undergoes the displacement of adjacent Te-Mo-Te layers to transition to the centrosymmetric $1T'$ phase, that is characterized by a monoclinic unit cell with $\alpha = 93.92°$. Both crystalline phases share a distorted intralayer structure in which the configuration of Mo centers surrounded by their neighboring Te atoms deviates substantially from the ideal octahedron of the trivial $1T$ state, presenting alternating shorter and longer Mo-Mo distances[9], see Fig. 1(a).

Furthermore, inversion symmetry breaking below $T_c$ results in the appearance of a characteristic low-frequency Raman-active interlayer shear phonon mode ($^1A_1 \approx 13$ cm$^{-1}$), see Fig. 1. This eminent $T_d \rightarrow 1T'$ phase-change signature and the possibility to control Weyl topologies through optical pumping of carriers have made the $T_d$ phase an attractive candidate to explore the effects of femtosecond (fs) optical excitation[10–15]. Time- and angle-resolved photoelectron spectroscopy (tr-ARPES) revealed an enhanced relaxation rate of photoexcited electrons when compared to the $1T'$ phase[10] as well as light induced band gap renormalization effects[15]. In a detailed investigation based on time-resolved single color transient reflectivity (tr-TR) and time-resolved second harmonic generation (tr-SHG) measurements, Zhang et al.[11,12] reported the discovery of an ultrafast photoinduced $T_d \rightarrow 1T'$ phase transition that proceeds in only $\sim 700$ fs. Moreover, a very recent time-resolved ultrafast electron diffraction (tr-UED) study by Qi et al.[14] provides evidence for the photoinduced suppression of the intralayer distortion in about 300 fs with a concomitant, albeit small, intralayer displacement of only a few picometers (pm). The later observation is consistent with the small interlayer displacement of ~ 8 pm observed by tr-UED following intense THz field excitation of the analogous compound, $T_d$-WTe$_2$[13]. The use of SHG in the study of single- to multi-layer 2D-TMDCs[16–20] has gained attraction owing to fact that the second order nonlinear susceptibility, $\chi^{(2)}$, vanishes in centrosymmetric systems. However, a discrepancy seems to exist between the conclusions drawn from the interpretation of structural data arising from tr-UED[13,14] experiments and those attained via tr-SHG measurements[11–13]. The latter appear to indicate the nearly disappearance of the non-centrosymmetric character of $T_d$-MTe$_2$ (M = Mo, W), and therefore a complete transformation to a $1T'$-like phase. It should be noticed that, unlike most commonly employed nonlinear optical crystals, the $\chi^{(2)}$ of 2D-TMDCs has shown to be strongly dependent on strain[16–19] and temperature[11,12,20], complicating the analysis of tr-SHG[11–14] signals, and thus justifying this apparent controversy.

In this letter, we show fluence-dependent time-resolved broadband transient reflectivity (tr-bb-TR) measurements carried out in a bulk single crystal of $T_d$-MoTe$_2$ at a base temperature of $T = 77$ K, i.e., well below $T_c$ as well as in the $1T'$ phase at room temperature, $T = 295$ K. Analyses of bb-TR and coherent $^1$A$_1$ Fourier intensity spectra as a function of the incident pump fluence ($F$) reveal a strong dependence on the probed photon energy ($E_{probe}$), deviations from the linear behaviour, as well as a progressive renormalization effect on electronic transitions that are susceptible to the coherent $^1$A$_1$ phonon displacement. Our results do not support the occurrence of the ultrafast photoinduced $T_d \rightarrow 1T'$ phase transition[11,12]. Instead, they suggest the progressive development of photoinduced interlayer strain[13] and/or the suppression of the intralayer distortion[14], acting as a competing electron-phonon relaxation pathway with increasing $F$.

In our tr-spectroscopic experiments, ≈ 100-fs pump pulses centered at a photon excitation energy of $E_{exc}$ = 1.2 eV (wavelength of $\lambda$ = 1030 nm) were implemented to drive the electron gas and the phonon bath out of equilibrium. Changes in the dielectric properties of the material arising from electronic population changes, hot-carrier energy relaxation processes, and coherent phonon oscillations were probed by time-delayed supercontinuum white light pulses with $E_{probe}$ spanning from ≈ 1.3 eV to 2.3 eV. Measurements were conducted using a quasi-collinear arrangement with an angle of 10º between the incident pump and probe beams, which carried out crossed polarizations. The probe beam and pump beam spot sizes at the crystal's surface were approximately 50 μm and 400 μm fwhm (full width at half maximum), respectively. High-quality tr-bb-TR data allowed the observation of carrier and phonon dynamics at gradually increasing $F$. The incident pump fluence was varied through the combination of a half-wave plate and a calcite polarizer. The repetition rate of laser system was 6 kHz and bb-TR spectra were obtained by modulating the pump beam with a mechanical chopper at 500 Hz. We minimized the collection time to avoid sample drift and excessive exposure to pump pulses and confirmed reversible sampling conditions up to the maximum $F \approx$ 5 mJ·cm$^{-2}$ presented in this work. 1$T'$-MoTe$_2$ crystals were grown in house following the recipe of Zeum et al.[9], and characterized via Raman and transport measurements[5,21].

Figure 1 illustrates some characteristic results obtained at $F$ = 1.90 mJ·cm$^{-2}$. The tr-bb-TR spectrum presented in Fig. 1(b) reveals several key features. The strong and long-period oscillation corresponds to the coherent $^1A_1$ interlayer phonon mode (≈ 13 cm$^{-1}$), whereas the weaker ones arise from the $^2A_1$ (≈ 77 cm$^{-1}$) and the $^5A_1$ (≈ 164 cm$^{-1}$) vibrational coherences[22]. In addition, there are clear spectral weight shifts and broadening that develop on the sub-picosecond (sub-ps) timescale following photoexcitation and are highlighted by the dashed rectangle in Fig.1 (b). The characteristic time for these spectral changes is in agreement with the 300-fs decay time observed for the suppression of the intralayer distortion[14] as well as the electron-electron and electron-phonon energy relaxation rate constants determined by tr-angle-resolved photoelectron spectroscopy in $T_d$-MoTe$_2$[10]. Moreover, very efficient electron-phonon coupling constants have been observed in several other 2D-TMDCs[23–27].

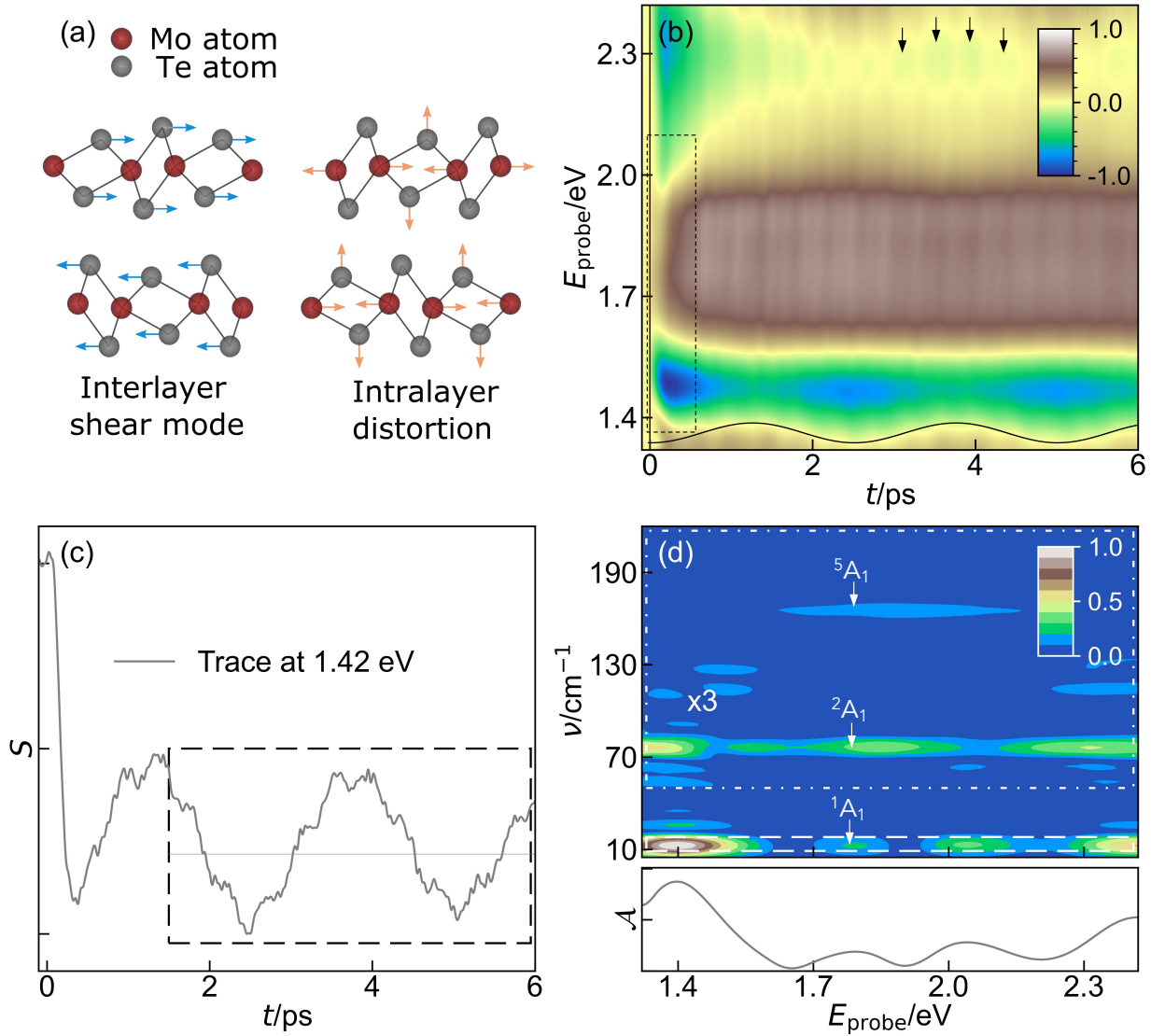

**FIG. 1.** Characteristic tr-bb-TR results obtained for $T_d$-MoTe$_2$ at 77 K and F = 1.9 mJ·cm$^{-2}$. (**a**) Illustration of the interlayer $^1$A$_1$ phonon mode (left) and the atomic displacements involved in the suppression of the intralayer distortion (right). (**b**) Chirp-corrected tr-bb-TR spectrum. The vertical solid line indicates time zero. The sinusoidal trace and arrows were introduced to guide the eye. The former illustrates the spectral amplitude modulation introduced by the $^1A_1 \sim 13$ cm$^{-1}$ phonon mode oscillation, which follows a cosine or displacive type of coherent phonon excitation. The latter pinpoints the period of the $^2A_1 \sim 77$ cm$^{-1}$ vibrational coherence. The dashed elongated rectangle highlights the observed spectral changes occurring at early times. (**c**) Temporal trace at $E_{probe}$ = 1.42 eV. The dashed rectangle indicates the portion of the trace employed to carry out further FFT analysis. (**d**) Fourier intensity map obtained from residuals as the one shown in panel **c**. The weaker FFT-amplitudes of the characteristic $^2A_1$ (77 cm$^{-1}$) and $^5A_2$ (164 cm$^{-1}$) phonons of the $T_d$ phase are also shown with their intensities scaled by a factor of 3 as indicated by the dash-dotted rectangle.

Figure 1(c) shows a slice of the tr-bb-TR spectrum at $E_{probe}$ = 1.42 eV. This time-trace was fitted to remove the electronic background signal and generate a residual (inset) for further fast Fourier transform (FFT) analysis. We limited our evaluations to $t > +1.5$ ps to circumvent the observed initial spectral changes and enable the automated fitting routine of large data sets. At $t > +1.5$ ps, electronic effects have reached a quasi-steady state, i.e., hot electrons have transferred most of their excess energy to the lattice via electron-electron and electron-phonon scattering processes within the conduction band. Therefore, time-dependent reflectivity changes at $t > +1.5$ ps are mostly governed by the oscillatory dynamics of coherently excited phonons. This procedure was applied at each value of $E_{probe}$, yielding the Fourier intensity map displayed in the top panel of Fig. 1(d). The bottom panel of Fig. 1(d) exhibits the values of the FFT-amplitude of the $^1A_1$ vibrational coherence ($\mathcal{A}$) as a function of $E_{probe}$, which were obtained through averaging along the small frequency interval specified by the dashed rectangle shown in the top panel of Fig. 1(d). We were able to determine the effects of increasing $F$ on the TR spectrum ($S$), $\mathcal{A}$ and the central frequency ($\nu_c$) of the coherent $^1A_1$ phonon. These findings are summarized in Fig. 2.

Different theories have been put forward to explain the mechanism for the generation of coherent $A_1$ phonons in simple semimetals and semiconductors, asserting the challenges involved in describing the driving force responsible for launching such coherent nuclear motion[28–37]. The sinusoidal curve in Fig. 1(b) pinpoints the phase of the $^1A_1$ vibrational coherence relative to time zero and indicates a cosine or displacive type of coherent phonon generation[28,35]. Figures 2(a) and 2(b) display the changes of $F$-normalized spectra ($S/F$ spectra) as a function of $F$ for the $T_d$ and $1T'$ phases, respectively. The bb-TR spectra were obtained by averaging data in the time domain between $t \approx +2.5$ ps and $+5$ps. This temporal interval corresponds approximately to one period of the $^1A_1$ phonon, see Fig. 1(c) and inset in Fig. 2(a), and this procedure was done to wash out the oscillatory effect caused by the $^1A_1$ and $^2A_1$ coherences on the TR signal. It is often expected for $S$ to scale approximately proportionally with $F$, or in other words to remain constant when normalized by $F$. This is indeed the case for the optical response of the $1T'$ phase, which is illustrated in Fig. 2(b). In contrast, the dielectric response of the $T_d$ state presents large deviations from linear behavior at $E_{probe} \approx 1.45$ eV. In addition, it is worth to mention that the TR spectra in Fig. 2(b) has been rescaled by a factor of 2 with respect to those shown in Fig. 2(a). These weaker TR signals can be explained by the higher baseline temperature of the $1T'$ phase, which leads to a higher population of thermal carriers, and therefore a relatively smaller change arising from the additionally photoexcited electron-hole pairs. The interpretation of the observed TR spectral changes is rather challenging since our $E_{probe}$ range involves high-energy interband transitions[38].

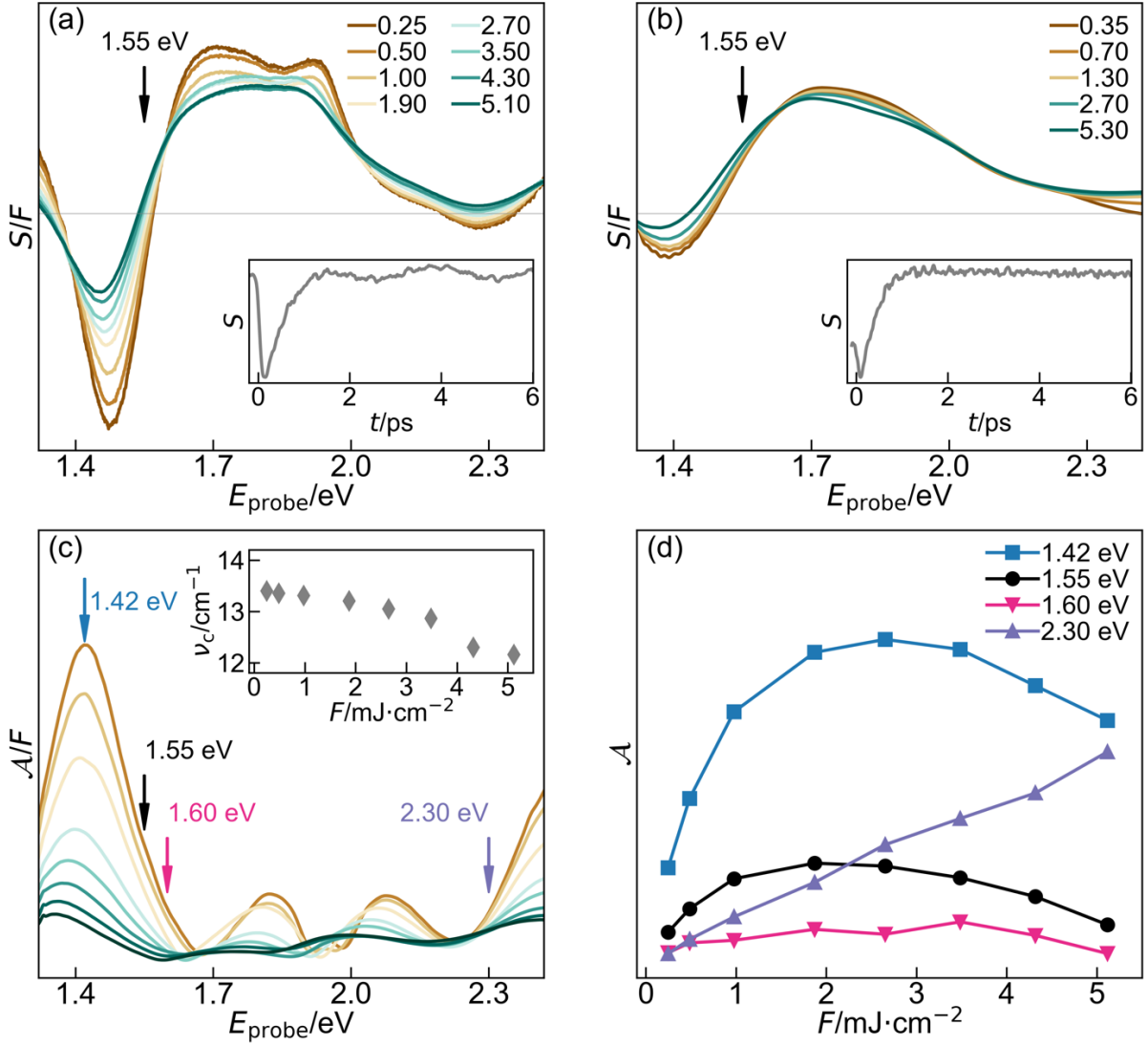

**FIG. 2.** (**a**), (**b**) Fluence-normalized transient spectra (S/F) for the $T_d$ (**a**) and $1T'$ (**b**) phases. The traces were time-averaged between $t \approx 2.5$ ps - 5ps (one $^1A_1$ phonon period) to wash out the effect of the $^1A_1$ phonon oscillation in the $T_d$ phase. The traces in panel **b** were scaled by a factor of 2. The insets show the characteristics temporal traces obtained at $E_{probe}$ = 1.55 eV (arrows) and $F$ = 1.90 mJ·cm² (**a**) and $F$ = 2.70 mJ· cm² (**b**). (**c**) Fluence-normalized $^1A_1$ Fourier intensity spectra ($\mathcal{A}/F$). The inset shows the $^1A_1$ phonon frequency determined as a function of $F$. These were obtained via fitting of time traces at $E_{probe} \approx 1.4$ eV using a sine function. The arrows indicate the values of $E_{probe}$ at which $\mathcal{A}$ is plotted on panel (**d**). (**d**) Fluence dependence of $\mathcal{A}$ at various $E_{probe}$.

Figure 2(c) exhibits the $F$-dependence of $F$-normalized $^1A_1$ Fourier intensity spectra ($\mathcal{A}/F$). The most noteworthy observation is the progressive redshift of $\mathcal{A}$ with increasing $F$. This observation suggests that the interband transitions that are dielectrically susceptible to the coherent $^1A_1$ phonon displacement experience a large degree of renormalization and they are likely the cause for the deviations observed in $S/F$. Figure 2(d) shows the $F$-dependence of $\mathcal{A}$ for some selected $E_{\text{probe}}$ values. Our results at $E_{\text{probe}} = 1.55$ eV are in reasonable agreement with those obtained by Zhang et at.[11,12]. However, our bb-data reveals no abrupt or threshold like behavior that could indicate the progression of a sub-ps photoinduced nonthermal $T_d \rightarrow 1T'$ phase transition[12]. Note that a phase transformation is expected to lead to the sudden disappearance of vibrational coherences[39]. As can be discerned by eye inspection of Fig. 3, there is a clear continuity of the amplitudes of the $^1A_1$ and $^2A_1$ coherent phonon modes even at $F = 5.1$ mJ·cm$^{-2}$. At this photoexcitation level and according to the energy balance, $\int_{77\,\text{K}}^{T_{l,m}} C(T)\ dT = \frac{F\,(1-R)\,M}{\rho\,\delta}$, the lattice should transiently reach a maximum temperature $T_{l,m} \approx 480$ K that is well above $T_c$ ($C$ is the heat capacity of $T_d$-MoTe$_2$[40], $\delta \approx 50$ nm is the optical penetration depth of the pump beam[41], $M = 351.1$ g·mol$^{-1}$ and $\rho = 7.78$ g·cm$^{-3}$, and $R \approx 0.40$ is the reflectivity at the crystal-vacuum interface[38]. We assumed that the latent heat for the $T_d \rightarrow 1T'$ phase transition is smaller than or of the order of the one for the transformation between $2H$ and $1T'$ polytypes (360 cal·mol$^{-1}$)[3]. This represents a small fraction ($\approx 0.1$ mJ·cm$^{-2}$) of the absorbed fluence ($\approx 3.1$ mJ·cm$^{-2}$) and can be then neglected. Hence, we found no evidence for an ultrafast photoinduced $T_d \rightarrow 1T'$ phase transition, neither nonthermal[12] nor thermal. Therefore, the observed TR changes and phonon dynamics are characteristic of a hotter $T_d$ phase.

Figure 3(a) also reveals that the overall decrease of $\mathcal{A}/F$ shown in Fig. 2(c) is not due to thermally induced decoherence affecting the damping constant of the coherent $^1A_1$ phonon. Note that the amplitude of the $^1A_1$ coherence at $E_{probe} \approx 1.4$ eV remains approximately constant within $t = +1.5$ ps and $+6.0$ ps. The increase of $F$, however, is found to lead to changes of the interlayer shear potential as evidenced by the strong softening of the $^1A_1$ phonon mode, see inset of Fig. 2(c).

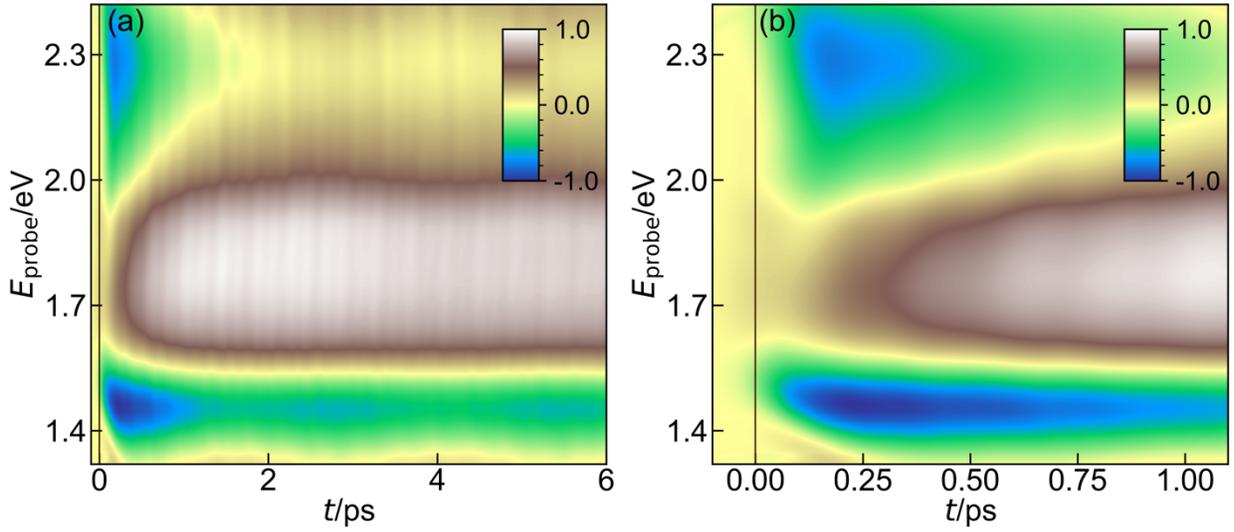

**FIG. 3.** (**a**) Chirp-corrected tr-bb-TR spectrum of $T_d$-MoTe$_2$ recorded at $T = 77$ K and $F = 5.1$ mJ·cm$^{-2}$. (**b**) Close-up view of the first picosecond following photoexcitation highlighting the smooth continuity of $^1A_1$ and $^2A_1$ coherences.

The success of known reversible ultrafast photoinduced phase transformations[24,39,42–50] rests on the ability of the sample to return to its original ground-state configuration before the next pair of pump and probe pulses excite and interrogate the system again. We would like to emphasize that the actual atomic rearrangements in crystalline systems capable of withstanding billions of photoexcitation cycles are well localized within the frame of an arbitrary unit cell. In contrast, the $T_d \rightarrow 1T'$ phase transition comprises the relative displacement of adjacent Te-Mo-Te layers. This process concerns many rigidly connected atoms that must respond collectively over the extent of a microscopic domain, and therefore it is unlikely to complete on the sub-ps timescale. Such a barrierless picture does not account for local variations in energy barriers and the degree of electronic delocalization. In fact, we believe that photoexcitation leads to a transiently hot $T_d$ state that dissipates the excess of energy into the bulk before having sufficient time to structurally transition to the $1T''$ phase.

Our results agree with the scenario of the photoinduced formation of a strained hot $T_d$ state with weaker cohesive interlayer forces, in which the suppression of the intralayer distortion[14], i.e. a localized atomic rearrangement, and the presence of domain walls in the *a-b* plane and along the stacking *c*-axis[51] may progressively facilitate a small shear displacement[13,14]. Photoinduced phase transitions are playing an increasingly important role in quantum materials and devices designed to confer properties on demand[52]. We hope that our discussion will bring a better understanding of the requirements for structural phenomena to be able to proceed on ultrafast time scales.

## Acknowledgements

We would like to thank Prof. Roberto Merlin (U. Michigan) and Dr. Ralph Ernstorfer (Fritz Haber Institute) for fruitful discussions and Prof. David Cory (U. Waterloo) for lending us the Oxford cryostat utilized in this work. G.S. acknowledge the support of the National Research Council of Canada, the Canada Foundation for Innovation, and Canada Research Chair program. A.W.T. acknowledges support from the US Army Research Office (W911NF-21-2-0136), Ontario Early Researcher Award (ER17-13-199), and the National Science and Engineering Research Council of Canada (RGPIN-2017-03815). This research was undertaken thanks in part to funding from the Canada First Research Excellence Fund. F.C., X. L. and Y.S. thank the support of the National

**Data availability**

The data that support the findings of this study are available from the corresponding author upon reasonable request.